%% file: paper.tex
\documentclass[fleqn,usenatbib]{mnras}

\usepackage{newtxtext,newtxmath}
\usepackage[T1]{fontenc}

\DeclareRobustCommand{\VAN}[3]{#2}
\let\VANthebibliography\thebibliography
\def\thebibliography{\DeclareRobustCommand{\VAN}[3]{##3}\VANthebibliography}

\usepackage{tikz}
\usetikzlibrary{calc}
\usepackage{booktabs}
\usepackage{array}
\newcolumntype{x}[1]{>{\centering\arraybackslash\hspace{0pt}}p{#1}}

\usepackage{graphicx}	% Including figure files
\usepackage{amsmath}	% Advanced maths commands

\usepackage{lipsum}

\graphicspath{{./figures/}}

\newif\ifedits
\editsfalse
\ifedits
    \newcommand{\edit}[1]{{\color{red} #1}}
    
\else
    \newcommand{\edit}[1]{#1}
    
\fi

\newcommand\JWST{{\it JWST}}

\defcitealias{Springel_Hernquist_2003}{SH03}
\defcitealias{Schaye_DallaVechhia_2008}{SDV08}
\defcitealias{Mannucci_2010}{M10}

\newcommand{\muZR}{ \mu_{\alpha}{\rm ZR}{} }

\newcommand{\betterorder}[1]{}

\title[Does the FMR evolve with redshift? I]{Does the Fundamental Metallicity Relation Evolve with Redshift? I: The Correlation Between Offsets from the Mass-Metallicity Relation and Star Formation Rate}
\author[Garcia et al.]{Alex M. Garcia$^{1}$\thanks{E-mail: alexgarcia@virginia.edu},
Paul Torrey$^{1}$, 
Sara Ellison$^{2}$,
Kathryn Grasha$^{4,5,6}$\thanks{ARC DECRA Fellow},
Lars Hernquist$^{3}$,
Henry R.M. Zovaro$^{4,5}$,\newauthor
Qian-Hui Chen$^{4,5}$,
Z.S. Hemler$^{7}$,
Lisa J. Kewley$^{3}$,
Erica J. Nelson$^{8}$,
Ruby J. Wright$^{9}$
\\~\\
$^{1}$Department of Astronomy, University of Virginia, Charlottesville, VA 22904, USA\\
$^{2}$Department of Physics \& Astronomy, University of Victoria, Finnerty Road, Victoria, British Columbia, V8P 1A1, Canada\\
$^{3}$Institute for Theory and Computation, Harvard-Smithsonian Center for Astrophysics, Cambridge, MA 02138, USA \\
$^{4}$Research School of Astronomy \& Astrophysics, Australian National University, Canberra, Australia, 2611 \\
$^{5}$ARC Centre of Excellence for All Sky Astrophysics in 3 Dimensions (ASTRO 3D), Australia\\
$^{6}$Visiting Fellow, Harvard-Smithsonian Center for Astrophysics, 60 Garden Street, Cambridge, MA 02138, USA\\
$^{7}$Department of Astrophysical Sciences, Princeton University, Peyton Hall, Princeton, NJ, 08544, USA \\
$^{8}$Department for Astrophysical and Planetary Science, University of Colorado, Boulder, CO 80309, USA\\
$^{9}$Department of Physics, University of Helsinki, Gustaf H{\"a}llstr{\"o}min katu 2, FI-00014 Helsinki, Finland
}

\date{Accepted XXX. Received YYY; in original form ZZZ}

\pubyear{\the\year}

\begin{document}
\label{firstpage}
\pagerange{\pageref{firstpage}--\pageref{lastpage}}
\maketitle

% Abstract of the paper
\begin{abstract}
The scatter about the mass-metallicity relation (MZR) has a correlation with the star formation rate (SFR) of galaxies.
The lack of evidence of evolution in correlated scatter at $z\!\lesssim\!2.5$ leads many to refer to the relationship between mass, metallicity, and SFR as the Fundamental Metallicity Relation (FMR).
Yet, recent high-redshift ($z>3$) \JWST{} observations have challenged the fundamental (i.e., redshift-invariant) nature of the FMR.
In this work, we show that the cosmological simulations Illustris, IllustrisTNG, and EAGLE all predict MZRs that exhibit scatter with a secondary dependence on SFR up to $z=8$.
We introduce the concept of a ``strong'' FMR, where the strength of correlated scatter does not evolve with time, and a ``weak'' FMR, where there {\it is} some time evolution.
We find that each simulation analysed has a \edit{statistically significant} weak FMR -- there is non-negligible evolution in the strength of the correlation with SFR.
Furthermore, we show that the scatter is reduced an additional $\sim$10-40\% at $z\gtrsim3$ when using a weak FMR, compared to assuming a strong FMR.
These results highlight the importance of avoiding coarse redshift binning when assessing the FMR.
\end{abstract}

% Select between one and six entries from the list of approved keywords.
% Don't make up new ones.

%%%%%%%%%%%%%%%%%%%%%%%%%%%%%%%%%%%%%%%%%%%%%%%%%%

%%%%%%%%%%%%%%%%% BODY OF PAPER %%%%%%%%%%%%%%%%%%

\section{Introduction}\label{sec:intro}

The metal content of galaxies provides key insights into galaxy evolution.
Stellar winds and supernovae explosions eject metals formed in stars into the interstellar medium (ISM).
Metals then mix via galactic winds \citep[e.g.,][]{Lacey_Fall_1985,Koeppen_1994} and turbulence \citep[e.g.,][]{Elmegreen_1999} within the disc while pristine gas accretion from the circumgalactic medium (CGM) and outflows dilute the metal content \citep[e.g.,][]{Somerville_Dave_2015}.
Thus, the metal content (metallicity) of the gas within a galaxy is sensitive to such processes, providing a window into the evolutionary processes within a galaxy \citep[][]{Dalcanton_2007,Kewley_2019,Maiolino_Mannucci_2019}.

Evidence for the sensitivity of metal content to the gas dynamics within a galaxy is perhaps most clearly seen within the relationship between the stellar mass of a galaxy and its gas-phase metallicity.
This mass-metallicity relationship (MZR) describes a relationship of increasing metal content in galaxies with increasing stellar mass \citep{Tremonti_2004,Lee_2006}.
At low stellar masses, the MZR relationship is well-described as a power-law, whereas at high masses ($\log[M_*/M_\odot] > 10.5$) the MZR plateaus \citep[e.g.,][]{Tremonti_2004,Zahid_2014,Blanc_2019}.
%The difference between these regimes is thought to be indicative of the efficacy of different processes in different mass galaxies, e.g. active galactic nuclei (AGN) feedback suppressing metal production in the highest mass galaxies \citep[e.g.,][]{DeRossi_2017}.
% Furthermore, it is now understood that the scatter about the MZR is not random. 
Furthermore, at a fixed stellar mass, low (high) metallicity galaxies have systematically elevated (depressed) gas masses (\citeauthor{Bothwell_2013} \citeyear{Bothwell_2013}; \citeauthor{Scholte_Saintonge_2023} \citeyear{Scholte_Saintonge_2023}) and SFRs (\citeauthor{Ellison_2008} \citeyear{Ellison_2008}; \citeauthor{Mannucci_2010} \citeyear{Mannucci_2010}).
The inverse relationship between a galaxy's metal content and SFR (or gas content) at a fixed stellar mass has been seen in the gas-phase in observations \citep[e.g.,][]{Lara_Lopez_2010,Bothwell_2016,Alsing_2024,Yang_2024} and simulations \citep[e.g.,][]{DeRossi_2017,Torrey_2018} as well as for stellar metallicities in simulations \citep[][]{DeRossi_2018,Fontanot_2021,Garcia_2024,Looser_2024} and recent observations \citep{Looser_2024}.
This secondary dependence on SFR and gas content is qualitatively well-described with basic competing physical drivers: 
(i) as new pristine gas is accreted onto a galaxy, it drives galaxies toward higher gas fractions, higher star formation rates (SFRs), and lower metallicities, while 
(ii) galaxies will persistently tend to consume gas and produce new metals, driving galaxies toward lower gas fractions, lower SFRs, and higher metallicities~\citep[e.g.,][]{Dave_2011,Dayal_2013,Lilly_2013,DeRossi_2015,Torrey_2018}. 
It is therefore expected that secondary dependence would remain present for galaxies across a wide redshift range given the ubiquity of these physical drivers.

At higher redshift the MZR has been seen to persist \citep[albeit with a lowered overall normalisation e.g.,][]{Savaglio_2005, Maiolino_2008, Zahid_2011, Langeroodi_2022} along with the secondary dependence on SFR \citep[e.g.,][]{Belli_2013,Salim_2015,Sanders_2018,Sanders_2021}.
Critically, it has been put forth that a single, redshift-invariant plane can be used to describe both the general evolution of the MZR as well as the secondary correlations \citep[][]{Mannucci_2010}.
This single surface/relation that can describe the metallicity of galaxies over a wide mass and redshift range is referred to as the fundamental metallicity relation (FMR).
\edit{
Despite the success of characterising galactic metallicities at $z\lesssim2.5$ ($\sim\!80\%$ of cosmic history), \cite{Mannucci_2010} report some evidence for deviations from the FMR at $z>3$. 
\JWST{} observations have recently corroborated the existence of deviations from the FMR
}\ignorespaces
at $z>3$ \citep{Heintz_2022,Curti_2023,Langeroodi_2023,Nakajima_2023}.

To remain truly redshift invariant, the FMR must capture two distinct features of the MZR simultaneously:
(i) the existence of a secondary relationship with SFR at fixed redshift, and 
(ii) the redshift evolution (or lack thereof) in the normalisation.
It is therefore possible that a change in either the MZR's secondary correlation with SFR or the redshift evolution of the normalisation of the MZR (or perhaps a combination thereof) may indicate FMR evolution.
Many of the previously mentioned studies investigating high-redshift galaxy populations apply a $z\sim0$ calibrated FMR to higher redshift data \citep[e.g.,][]{Mannucci_2010,Wuyts_2012,Belli_2013,Sanders_2021,Curti_2023,Langeroodi_2023,Nakajima_2023}.
However, it is unclear how to effectively decouple (and subsequently interpret) the observed evolution at high redshift in these frameworks.
Some work has been done up to this point observationally looking at higher redshifts independently to specifically isolate the scatter about the MZR \citep[e.g.,][]{Salim_2015,Sanders_2015,Sanders_2018,Li_2023,Pistis_2023}.
These works find that there may be some evolution within the scatter about the MZR at intermediate redshifts (\citeauthor{Pistis_2023} \citeyear{Pistis_2023} suggest potentially as low at $z\sim0.63$).
Yet, there are comparatively few simulations results on a systematic examinations on the strength of the secondary dependence on gas content and/or SFR at individual redshifts.

In this work, we investigate the redshift evolution of the MZR's secondary dependence on SFR from the perspective of the cosmological simulations Illustris, IllustrisTNG, and EAGLE.
The rest of the paper is as follows: 
In \S\ref{sec:methods} we describe the simulations we use, our galaxy selection criteria, and summarize definitions of the FMR.
In \S\ref{sec:results} we present the redshift evolution of the FMR as found in simulations.
In \S\ref{sec:discussion} we quantify the impact of the new framework on the scatter about the MZR, discuss the advantages and challenges in the new framework, and then discuss potential impacts of the physical models.
Finally, in \S\ref{sec:conclusion} we present our conclusions.

\section{Methods}
\label{sec:methods}

We use the Illustris, IllustrisTNG, and EAGLE cosmological simulations to investigate the dependence of the gas-phase metallicity on stellar mass and star formation.
Each of these simulations has a sub-grid ISM pressurisation model, which creates ``smooth'' stellar feedback. 
We believe that generic results from all three of these simulations should constitute a fair sampling of predictions from sub-grid ISM pressurisation models owing to the appreciably different physical implementations.

Here we briefly describe each of the simulations from this analysis, the galaxy selection criteria we employ, and present a new framework for interpreting the \citeauthor{Mannucci_2010} (\citeyear{Mannucci_2010}; hereafter \citetalias{Mannucci_2010}) FMR projection.
All measurements are reported in physical units.
% We highlight the differences between the simulation models in Table~\ref{tab:simulation_details}.
% Many of the methods here follow from \cite{Garcia_2024}.

\subsection{Illustris}\label{subsec:Illustris}

The original Illustris suite of cosmological simulations \citep[][]{Vogelsberger_2013,Vogelsberger_2014a,Vogelsberger_2014b,Genel_2014,Torrey_2014} was run with the moving-mesh code {\sc arepo} \citep[][]{Springel_2010}.
The Illustris model accounts for many important astrophysical processes, including gravity, hydrodynamics, star formation/stellar evolution, chemical enrichment, radiative cooling and heating of the ISM, stellar feedback, black hole growth, and AGN feedback.
The unresolved star forming ISM uses the \citeauthor{Springel_Hernquist_2003} (\citeyear{Springel_Hernquist_2003}) equation of state, wherein new star particles are created from regions of dense ($n_{\rm H} > 0.13$~cm$^{-3}$) gas.
The masses of the stars within the star particle are drawn from a \cite{Chabrier_2003} initial mass function (IMF) and metallicities are adopted from the ISM where they are born.
As the stars evolve, they eventually return their mass and metals back into the ISM.
The stellar mass return and yields used allow for the direct simulation of time-dependent return and heavy metal enrichment, explicitly tracking nine different chemical species (H, He, C, N, O, Ne, Mg, Si, and Fe).

The Illustris suite consists of a single volume of size (106.5 Mpc)$^3$ at three different resolutions.
The three resolutions are as follows: Illustris-1 ($2\times1820^3$ particles), Illustris-2 ($2\times910^3$ particles), and Illustris-3 ($2\times455^3$ particles).
We use Illustris-1, the highest resolution run, which is hereafter we refer to synonymously with Illustris itself.

\subsection{IllustrisTNG}\label{subsec:TNG}

IllustrisTNG \citep[The Next Generation;][hereafter TNG]{Marinacci_2018,Naiman_2018,Nelson_2018,Pillepich_2018b,Springel_2018,Pillepich_2019, Nelson_2019a, Nelson_2019b} is the successor to the original Illustris simulations, alleviating some of the deficiencies of and updating the original Illustris model.
As such, the Illustris and TNG models are similar, yet have an appreciably different physical implementation (see \citeauthor{Weinberger_2017} \citeyear{Weinberger_2017}; \citeauthor{Pillepich_2018a} \citeyear{Pillepich_2018a}, for a complete list of differences between the models).
A critical difference between the Illustris and TNG models for the context of this work is TNG's implementation of redshift-scaling winds.
The TNG model employs a wind velocity floor not present in the original Illustris model in order to prevent low mass haloes from having unphysically large mass loading factors.
Consequently, low redshift star formation is suppressed in the TNG model.
TNG implements the same equation of state for the dense star forming ISM as Illustris \citep{Springel_Hernquist_2003}.
As in Illustris, new star particles are created from dense gas using the \cite{Chabrier_2003} IMF.
Furthermore, TNG tracks the same nine chemical species as Illustris, while also following a tenth ``other metals'' as a proxy for metals not explicitly monitored.

TNG consists of three different volumes each with their own sub-resolution runs: TNG50 (51.7 Mpc)$^3$, TNG100 (110.7 Mpc)$^3$, and TNG300 (302.6 Mpc)$^3$.
In this work, we will use the highest resolution TNG100 run (TNG100-1; hereafter used synonymously with TNG), with $2\times1820^3$ particles, as a comparable volume and resolution to the original Illustris.

\subsection{EAGLE}\label{subsec:EAGLE}

Unlike Illustris and TNG, ``Evolution and Assembly of GaLaxies and their Environment'' \citep[EAGLE,][]{Crain_2015, Schaye_2015, McAlpine_2016} employs a heavily modified version of the smoothed particle hydrodynamics (SPH) code {\sc gadget-3} (\citeauthor{Springel_2005} \citeyear{Springel_2005}; {\sc anarchy}, see \citeauthor{Schaye_2015} \citeyear{Schaye_2015} Appendix A).
EAGLE includes many of the same baryonic processes (star-formation, chemical enrichment, radiative cooling and heating, etc) as Illustris and TNG.
The dense (unresolved) ISM in EAGLE is also treated with a sub-grid equation of state (\citeauthor{Schaye_DallaVechhia_2008}~\citeyear{Schaye_DallaVechhia_2008}; hereafter, \citetalias{Schaye_DallaVechhia_2008}), much like that of \citetalias{Springel_Hernquist_2003}.
The \citetalias{Schaye_DallaVechhia_2008} prescription forms stars according to a \cite{Chabrier_2003} IMF from the dense ISM gas.
The density threshold for star formation is given by the metallicity-dependent transition from atomic to molecular gas computed by \cite{Schaye_2004} with an additional temperature-dependent criterion \citep[][]{Schaye_2015}.
Stellar populations evolve according to the \cite{Wiersma_2009b} evolutionary model and eventually return their mass and metals back into the ISM.
EAGLE explicitly tracks eleven different chemical species (H, He, C, N, O, Ne, Mg, Si, S, Ca, and Fe).

The full EAGLE suite is comprised of several simulations ranging from size $(12~{\rm Mpc})^3$ to $(100~{\rm Mpc})^3$.
We use data products at an intermediate resolution ($2\times1504^3$ particles) run with a box-size of ($100~{\rm Mpc})^3$ referred to as {\sc RefL0100N1504} (hereafter simply EAGLE) as a fair comparison to the selected Illustris and TNG runs.

\subsection{Galaxy selection}\label{subsec:galaxy_selection}

All three simulations in this work select gravitationally-bound substructures using {\sc subfind} \citep[][]{Springel_2001,Dolag_2009}, which identifies self-bound collections of particles from within friends-of-friends groups (\citeauthor{Davis_1985} \citeyear{Davis_1985}).
We limit our analysis to central galaxies that we consider `well-resolved' (i.e., containing $\sim$100 star particles and $\sim$500 gas particles), thus we restrict the sample to galaxies with stellar mass $\log(M_*~[M_\odot]) > 8.0$ and gas mass $\log(M_{\rm gas}~[M_\odot]) > 8.5$.
We place an upper stellar mass limit of $\log(M_*~[M_\odot]) > 12.0$\footnote{The upper mass limit does not exclude any galaxies for most redshifts}.
Following from a number of previous works (see, e.g., \citeauthor{Donnari_2019} \citeyear{Donnari_2019}; \citeauthor{Nelson_2021} \citeyear{Nelson_2021}; \citeauthor{Hemler_2021} \citeyear{Hemler_2021}; \citeauthor{Garcia_2023a} \citeyear{Garcia_2023a}), we exclude quiescent galaxies by defining a specific star formation main sequence (sSFMS).
We do so by fitting a linear-least squares regression to the median sSFR-M$_*$ relation with stellar mass $\log(M_*~[M_\odot]) < 10.2$ in mass bins of 0.2 dex.
The sSFMS above $10.2 \log M_\odot$ is extrapolated from the regression.
Galaxies that fall greater than 0.5 dex below the sSFMS are not included in our sample.
As we show in \citeauthor{Garcia_2024} (\citeyear{Garcia_2024}; that paper's Appendix B), our key results (using stellar metallicities) are {\it not} sensitive to our sample selection.
We obtain the same result here in the gas-phase: our key results are qualitatively unchanged by the same variations as \cite{Garcia_2024} in selection criteria (see Appendix~\ref{appendix:sSFMS_cuts}).

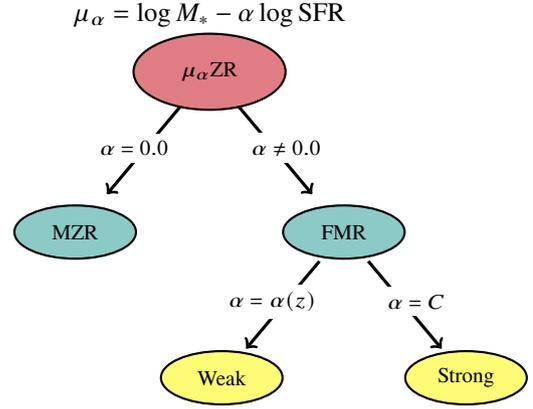
\begin{figure}
    \centering
    \input{figures/Figure1.tikz}
    \caption{{\bf Decision tree for the $\muZR$, see Section~\ref{subsec:FMR_param} for full details.} This shows the different relationships that can be included under the umbrella $\mu_{\alpha}$ metallicity relation ($\muZR$; see Equation~\ref{eqn:mu_M10}). First is the traditional MZR where $\alpha=0.0$ and second is the FMR where $\alpha\neq0.0$.
    The FMR can be further broken into two categories: strong and weak depending on if $\alpha$ varies as a function of redshift (weak) or not (strong).}
    \label{fig:general_muZR}
\end{figure}

As metallicity measurements are typically limited to star forming regions in observations \citep[][]{Kewley_Ellison_2008,Kewley_2019}, all of the analysis of gas-phase metallicities presented here is based only on star-forming gas (as defined in Section~\ref{subsec:Illustris} for Illustris/TNG and Section~\ref{subsec:EAGLE} for EAGLE).

\subsection{Definitions of the FMR}\label{subsec:FMR_param}

\citetalias{Mannucci_2010} propose that the 3D relationship between stellar mass, gas-phase metallicity, and star formation rate (SFR) can be projected into 2D using a linear combination of the stellar mass and star formation:
\begin{equation}
    \mu_{\alpha} = \log M_* - \alpha \log {\rm SFR}~,
    \label{eqn:mu_M10}
\end{equation}
where $\alpha$ is a free parameter that ranges from 0 to 1\ignorespaces
\footnote{
\edit{In reality, the correlated scatter about the MZR has some non-negligible mass dependence.
In fact, the correlation with SFR has been seen to weaken, or even invert, at high stellar mass \citep[e.g.,][]{Yates_2012,Alsing_2024}.}
Parameterisations exist \edit{that account for this mass dependence exist} \citep[e.g.,][]{Curti_2020}.
We opt to not present other forms of the FMR in this work as a exercise on the extent to which the \citetalias{Mannucci_2010} projection can describe the of scatter at fixed redshift \edit{(see further discussion in Section~\ref{subsec:evaluting_FMR})}.}.
The free parameter $\alpha$ holds all the diagnostic power on the strength of the MZR's secondary dependence with SFR.
By varying $\alpha$, the distribution of galaxies in $\mu_{\alpha}$-metallicity space varies.
We define a $\mu_{\alpha}$-metallicity relation ($\muZR$) for each $\alpha$ as a linear-least squares regression\ignorespaces
\footnote{\citetalias{Mannucci_2010} use a fourth-order polynomial for fitting.  This practice is inconsistent in the literature with many \citep[e.g.,][]{Andrews_Martini_2013} considering a linear regression. We show that using a fourth-order polynomial instead of a linear regression does not significantly alter our $\alpha_{\rm min}$ determination in Appendix~\ref{appendix:polynomial}.} of the data.
We compute the $\muZR$ for $\alpha=0.0~{\rm to}~1.0$ in steps of $0.01$ and obtain the residuals about each regression.
The projection that yields the minimum scatter in the residuals (smallest standard deviation) is deemed the best fit.
The $\alpha$ value associated with this minimum scatter projection is henceforth referred to as $\alpha_{\rm min}$.
We define an uncertainty on $\alpha_{\rm min}$ by assuming that a projection that has scatter within 5\% of the minimum value is a plausible candidate for the true $\alpha_{\rm min}$ \citep[following from][]{Garcia_2024}.

$\alpha_{\rm min}$ physically represents the direction to project the 3D mass-metallicity-SFR ($M_*\!-\!Z_{\rm gas}\!-\!{\rm SFR}$) space into a minimum scatter distribution in 2D $\mu_{\alpha}\!-\!Z$ space.
% Investigation of the FMR in the \citetalias{Mannucci_2010} framework is ``Can the 2D projection explain both:
% (i) the scatter about the MZR, and
% (ii) the evolution of normalisation in the MZR through time.''
Thus, the $\muZR$ is the relation of merit in the 2D projection of the $M_*-Z_{\rm gas}-{\rm SFR}$ relation.
There are two outcomes, either
(i) $\alpha_{\rm min}=0.0$, wherein the canonical MZR is recovered, or
(ii) $\alpha_{\rm min}\neq0.0$, wherein an FMR is recovered.
In this way, the $\muZR{}$ can be thought of as a superset of relations containing the MZR, the strong FMR, and the weak FMR (relationships illustrated in Figure~\ref{fig:general_muZR}).
Framing the FMR in this way underscores the decisions required in establishing the FMR.
Previous studies have been somewhat restrictive in regards to these decisions.
We therefore highlight the need to take a deliberate approach to our definitions to build a framework by which potential redshift evolution can be assessed.

Traditionally (as in, e.g., \citetalias{Mannucci_2010}), the FMR is defined by determining $\alpha_{\rm min}$ at $z=0$.
This value has been seen to be roughly constant at $z\lesssim2.5$ (e.g., \citetalias{Mannucci_2010}; \citeauthor{Andrews_Martini_2013} \citeyear{Andrews_Martini_2013}).
We henceforth refer to the idea that $\alpha_{\rm min}$ does not vary as a function of redshift as the strong FMR.
A single $\alpha_{\rm min}$ can describe both the MZR's secondary dependence and its normalisation evolution in the strong FMR.
In this work, we investigate the claim that $\alpha_{\rm min}$ is constant over time by identifying the $\alpha_{\rm min}$ value that minimizes scatter at each redshift independently.
This procedure allows $\alpha_{\rm min}$ to (potentially) vary as a function of redshift.
Here we introduce the concept of a ``weak'' FMR.
We define the weak FMR as a counterpoint to the strong FMR: that $\alpha_{\rm min}\neq0$, but $\alpha_{\rm min}$ is {\it not} constant as a function of redshift (see illustrated relationship in Figure~\ref{fig:general_muZR}).

There are actually more parameters beyond $\alpha_{\rm min}$ that the FMR is defined by: the parameters of the regression (in our case slope and intercept).
These additional parameters add complexity to the interpretation of the evolution.
Regressions are inherently linked to the $\alpha_{\rm min}$ determination, yet the parameters of the fit can have a profound impact on interpretation of FMR evolution irrespective of $\alpha_{\rm min}$ variations.
The impact of these parameters is beyond the scope of this work since we only examine each redshift bin independently here and the effect of the regression parameters is only felt when comparing different redshift bins.
We do, however, address the impact of these parameters in a companion work (Garcia et al. in prep).

\section{Results}
\label{sec:results}

\input{tables/table1}

\begin{figure*}
    \centering
    \includegraphics[width=\linewidth]{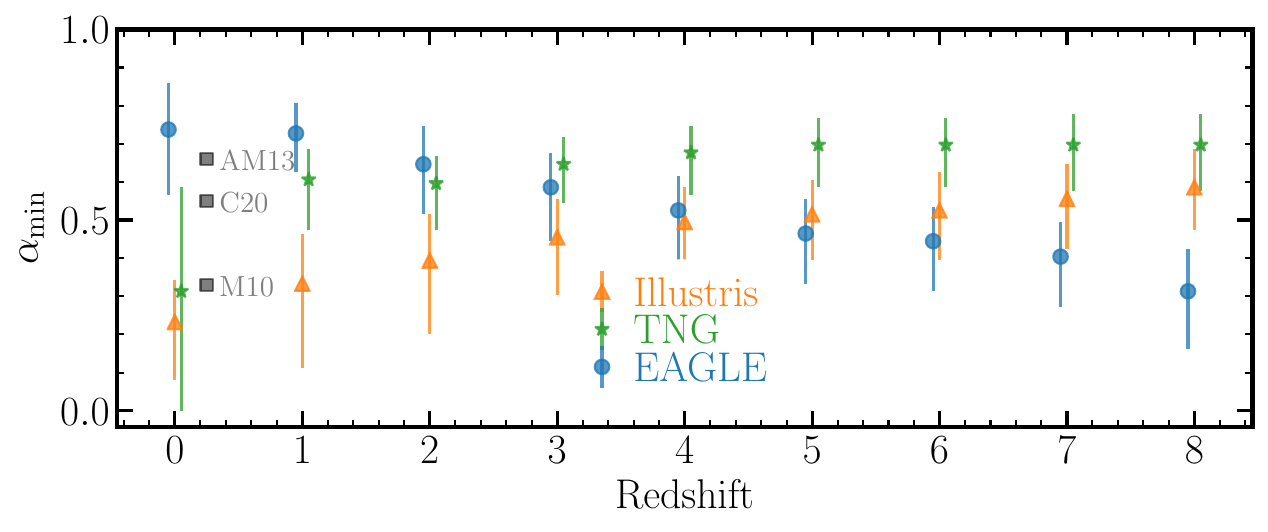}
    \caption{ {\bf $\alpha_{\rm min}$ values as a function of redshift in Illustris, TNG, and EAGLE.} $\alpha_{\rm min}$ values as a function of redshift are plotted as orange triangles, green stars, and blue circles for Illustris, TNG, and EAGLE, respectively. The errorbars here are obtained by finding $\alpha$ values that reduce the scatter to within 5\% that of the minimized scatter. The gray squares are observational values of $\alpha_{\rm min}$ from \citetalias{Mannucci_2010}, \citeauthor{Andrews_Martini_2013} (\citeyear{Andrews_Martini_2013}), and \citeauthor{Curti_2020} (\citeyear{Curti_2020}) determined at $z\approx0$ via SDSS (offset from $z=0$ for aesthetic purposes). }
    \label{fig:alpha_redshift}
\end{figure*}

\subsection{Does \texorpdfstring{$\alpha_{\rm min}$}{alpha min} vary as a function of redshift?}\label{subsec:alpha_redshift}

We use the best-fit $\alpha_{\rm min}$ values derived as a function of redshift to evaluate whether the scatter about the MZR evolves significantly with redshift.
We find that $\alpha_{\rm min} \neq 0$ at all redshifts in each of the three simulations (Figure~\ref{fig:alpha_redshift}).
Based on the first step of the decision tree in Figure~\ref{fig:general_muZR}, the non-zero $\alpha_{\rm min}$ values show there is an FMR in each simulation.
The secondary dependence on SFR is present, at least to some extent, within the scatter of all the MZRs analysed here.
It should be noted, however, that the uncertainty\footnote{Uncertainties on $\alpha_{\rm min}$ correspond to the uncertainty in the minimum dispersion (see Section~\ref{subsec:FMR_param} for definition)} on the TNG $z=0$ $\alpha_{\rm min}$ value does include $\alpha=0$.
This implies a somewhat weak dependence on SFR at this redshift.
In \cite{Garcia_2024}, we attribute a lack of a relationship at $z=0$ in TNG to the redshift scaling of winds within the TNG model\footnote{We show this in \cite{Garcia_2024} for {\it stellar} metallicities. That work also demonstrates that stellar and gas-phase metallicities are related to each other (see Section 4.1 of that work). Therefore, the same physical mechanism suppressing the correlated scatter for stellar metallicities is likely what is suppressing $\alpha_{\rm min}$ for the gas-phase.}.
Briefly, the effect of adding winds that change with redshift suppresses low redshift star formation and increases the efficiency of high redshift stellar feedback compared to the Illustris model (see \citeauthor{Pillepich_2018a} \citeyear{Pillepich_2018a}).
It is therefore likely that the suppressed low redshift star formation causes the large uncertainty on $\alpha_{\rm min}$ at $z=0$.
% The lack of evolution in $\alpha_{\rm min}$ at higher redshifts may be a feature of the increased efficacy of stellar feedback at these higher redshifts.
As such, features of $\alpha_{\rm min}$ are sensitive to details of the wind implementation/strength prescribed by the model on which it is built (see Section~\ref{subsec:subgrid_physics} for further discussion).

Overplotted on Figure~\ref{fig:alpha_redshift} (gray squares) are three observationally determined values of $\alpha_{\rm min}$ from \citetalias{Mannucci_2010} (0.32), \citeauthor{Andrews_Martini_2013} (\citeyear{Andrews_Martini_2013}; 0.66), and \citeauthor{Curti_2020} (\citeyear{Curti_2020}; 0.55).
Each of these values was determined using SDSS galaxies at $z\approx0$ (offset horizontally for clarity).
Deviations in the observational values are attributed primarily to: (i) different metallicity calibrations, (ii) using individual galaxies versus galaxy stacks (as in \citeauthor{Andrews_Martini_2013} \citeyear{Andrews_Martini_2013}), and (iii) selection biases towards higher star forming galaxies.
Simulations are not directly affected by metallicity calibrations in the same way as observations.
The sample selection criteria (outlined in Section~\ref{subsec:galaxy_selection}) should help mitigate the effect of selection function biases of observations.
Though we select star-forming galaxies, we do not just select the highest star forming galaxies.
In spite of these potential differences, it is worth noting that the \citetalias{Mannucci_2010} $\alpha_{\rm min}$ value agrees fairly well with the TNG and Illustris derived values at $z=0$. 
Although the uncertainty on the TNG $\alpha_{\rm min}$ is significant enough to include the \cite{Curti_2020} value by a factor of $\sim\!1.5$ times higher.
Similarly, the \cite{Andrews_Martini_2013} value of $0.66$ agrees fairly well with the derived value from EAGLE at $z=0$, though we caution that this analysis was done with galaxy stacks whereas we us individual galaxies here.

% The obtained $\alpha_{\rm min}$ values at $z=0-8$ are shown in Figure~\ref{fig:alpha_redshift} and presented in Table~\ref{tab:alphas}.
Furthermore, we find that $\alpha_{\rm min}$ values show some level of redshift evolution in all three simulations (Figure~\ref{fig:alpha_redshift} and Table~\ref{tab:alphas}).
Interestingly, each simulation has qualitatively different redshift evolution.
TNG $\alpha_{\rm min}$ values vary significantly from $z=0$ to $z=1$ but then level off, the Illustris $\alpha_{\rm min}$ values increase monotonically with redshift, and the EAGLE values decrease monotonically as a function of redshift.

\edit{
We conduct a one-sample $t$-test to validate this apparent redshift evolution in each simulation.
% The one-sample $t$-test assess whether the mean value of $\alpha_{\rm min}$ significantly differs from its value at $z=0$.
The null hypothesis is that the mean $\alpha_{\rm min}$ is equal to $\alpha_{\rm min}$ at $z=0$ (i.e., there is {\it not} redshift evolution).
Given the uncertainty associated with each $\alpha_{\rm min}$, we compute the sample mean by weighting each $\alpha_{\rm min}$ by the reciprocal of its squared uncertainty\ignorespaces
\footnote{We note that we make the simplifying assumption of symmetric uncertainty by defining the offset as the average of the upper and lower offsets.
We additionally verify that instead choosing the magnitude of either the upper or lower offsets does change the result.
}.
Additionally, we normalize the $t$-statistic by the estimated error on the mean (the reciprocal sum of squared weights).
We find that $t$-statistics are $-6.23$, $13.34$, and $7.04$ in Illustris, TNG, and EAGLE (respectively).
These correspond to $p$-values of $2.5\times10^{-4}$ for Illustris, $9.5\times10^{-7}$ for TNG, and $1.0\times10^{-4}$ EAGLE.
We therefore reject the null hypothesis at a significance level of $p=0.05$ in each simulation, indicating statistically significant redshift evolution of $\alpha_{\rm min}$.}
From the decision tree of Figure~\ref{fig:general_muZR}, the strong FMR is ruled out in favour of a weak FMR for each individual simulation.
% Redshift evolution of the derived $\alpha_{\rm min}$ values may be indicative of evolution in the driving forces of the scatter about MZR.
\edit{
It is interesting to note that if we test instead using the $z=1$ value of $\alpha_{\rm min}$ in TNG we obtain a $t$-statistic of $1.64$ ($p$-value of $0.139$).
We could not reject the null hypothesis of no evolution of the FMR at the 0.05 significance level in that case (see above discussion on redshift-scaling winds in TNG model).
}

Our result in EAGLE indicating significant redshift evolution seemingly contradicts a previous study finding the FMR is in place and does not evolve out to $z\approx5$ in EAGLE \citep[][]{DeRossi_2017}.
There is a subtle difference in the analysis between the two works, however: \cite{DeRossi_2017} do not parameterise the FMR to test $\alpha_{\rm min}$ variations.
They qualitatively examine the secondary dependence within the MZR and show that a $M_*\!-\!Z_{\rm gas}\!-\!{\rm SFR}$ relation exists at $z=0-5$ (i.e., there is at least a weak FMR over these redshift ranges).
% This is in agreement with our results here.
We find that an $M_*\!-\!Z_{\rm gas}\!-\!{\rm SFR}$ relation at $z=0-5$ exists in EAGLE via non-zero $\alpha_{\rm min}$ values (i.e., there is at least a weak FMR over these redshift ranges), consistent with \cite{DeRossi_2017}.
Despite the persistence of the $M_*\!-\!Z_{\rm gas}\!-\!{\rm SFR}$ relation, we confirm that there is a weak in EAGLE by using the \citetalias{Mannucci_2010} projection of the FMR.
It should be noted that the uncertainty of the $z=0$ and $z=5$ values do overlap in EAGLE.
The subtly of the redshift evolution may therefore be difficult to detect without fitting each redshift independently.

\subsection{Scatter assuming different FMRs}\label{subsubsec:discussion_alphamin}

\begin{figure*}
    \centering
    \includegraphics[width=\linewidth]{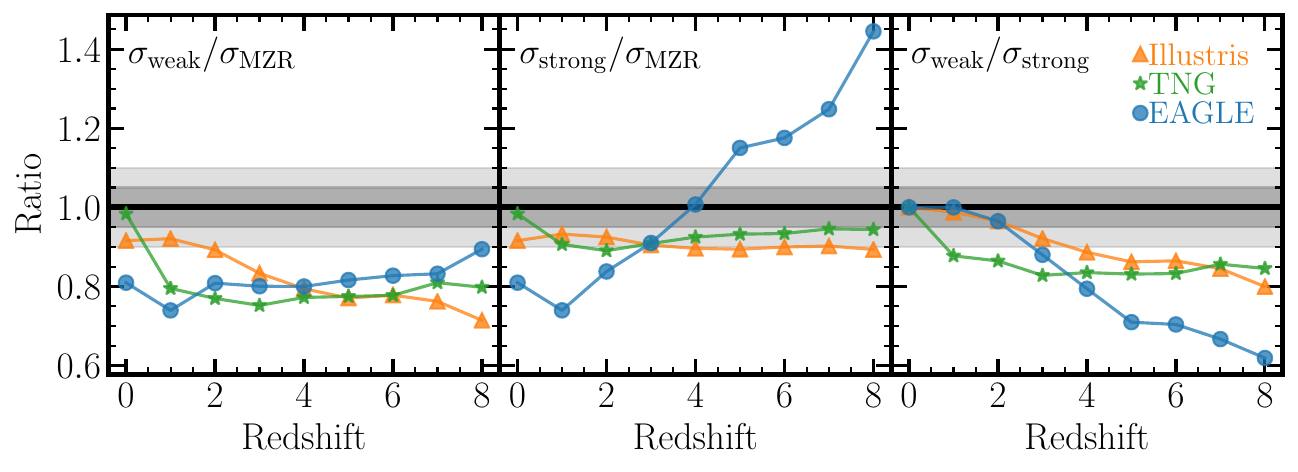}
    \caption{{\bf Reduction in scatter for weak FMR versus MZR, strong FMR versus MZR, and weak FMR versus strong FMR.} {\it Left:} the scatter about the FMR by fitting $\alpha_{\rm min}$ at each redshift individually ($\sigma_{\rm weak}$) divided by the scatter in the MZR at each redshift ($\sigma_{\rm MZR}$) as a function of redshift. The dark and light gray shaded regions (in all panels) represent 5\% and 10\% variations, respectively, of each ratio. {\it Centre:} Same as left, but now the numerator is the scatter about the FMR evaluated in each redshift bin with a $z=0$ calibrated $\alpha_{\rm min}$ ($\sigma_{\rm strong}$). {\it Right:} Previous two panels divided by each other, the reduction in scatter in the relationship by determining $\alpha_{\rm min}$ at each redshift independently ($\sigma_{\rm weak}$) divided by using the $z=0$ $\alpha_{\rm min}$ value ($\sigma_{\rm strong}$).}
    \label{fig:scatter_reduction}
\end{figure*}

The derived $\alpha_{\rm min}$ values show that Illustris, TNG, and EAGLE all have weak FMRs.
We now examine how the impact of marginalising over variations in $\alpha_{\rm min}$ when assuming a strong FMR.
Specifically, we want to quantify how the scatter changes when assuming a strong versus weak FMR.
This will provide a quantitative metric for assessing how important considerations for an evolving FMR are.
If the scatter were to remain unchanged, or change only marginally, the need for a weak FMR would be minimal.

To this end, we define three different ratios for quantitatively evaluating the importance of using a weak FMR.
We consider ratios of the scatters about:
(i) the weak FMR compared to the MZR,
(ii) the strong FMR compared to the MZR, and
(iii) the weak FMR compared to the strong FMR.
Figure~\ref{fig:scatter_reduction} illustrates these three different ratios evaluated at each redshift for Illustris (orange diamonds), TNG (green stars), and EAGLE (blue circles).
\edit{
We note that we calculate scatter about the MZR in the same way as we used to determine $\alpha_{\rm min}$ in the previous section (i.e., using the projection from Equation~\ref{eqn:mu_M10}) in the following discussion.
}

The left panel of Figure~\ref{fig:scatter_reduction} shows the standard deviation of the residuals (henceforth, scatter) about each redshift's weak FMR ($\alpha_{\rm min}$ determined at that redshift) normalised by the scatter about the MZR ($\alpha_{\rm min}=0$) as a function of redshift ($\sigma_{\rm weak}/\sigma_{\rm MZR}$).
We find for all redshifts across the three simulations that the weak FMR reduces the scatter by $\sim\!\!10-30\%$ compared to the MZR.
The exception is TNG at $z=0$ having scatter reduction of less than $5\%$ -- falling within the nominal uncertainty on $\alpha_{\rm min}$ and implying there is functionally no difference between the scatter of the MZR and FMR at this redshift.
This $z=0$ TNG exception was discussed previously in the context of the $\alpha_{\rm min}$ value (see Section~\ref{subsec:alpha_redshift}) and the lack of a relation was attributed to the redshift scaling winds in the TNG model (see \citeauthor{Pillepich_2018a} \citeyear{Pillepich_2018a}).
The scatter reduction is roughly constant as a function of redshift in both TNG and EAGLE at around $\sim\!20\%$ (barring the aforementioned TNG exception).
Scatter reduction in Illustris ranges from $\lesssim\!10\%$ at $z=0$ to nearly 30\% at $z=8$.

The middle panel of Figure~\ref{fig:scatter_reduction} shows the scatter at each redshift assuming a strong FMR compared to that of the MZR at that redshift ($\sigma_{\rm strong}/\sigma_{\rm MZR}$).
We define the strong FMR fit analogously to observations: we apply a $z=0$ determined $\alpha_{\rm min}$ value to all redshifts.
In TNG, we find a similar trend to that of $\sigma_{\rm weak}/\sigma_{\rm MZR}$: a roughly constant scatter reduction as a function of redshift, albeit at a reduced value of $\sim\!\!5-10\%$ (see previous discussion about the exception at $z=0$).
The scatter reduction in Illustris is similarly constant around $10\%$.
Evidently, the redshift evolution in Illustris seen previously with $\sigma_{\rm weak}/\sigma_{\rm MZR}$ disappears when assuming a strong FMR.
$\sigma_{\rm strong}/\sigma_{\rm MZR}$ actually {\it increases} nearly monotonically in EAGLE as a function of redshift: the strong FMR fit on the low redshift bins is significantly better than the highest redshifts.
Remarkably, assuming a strong FMR actually begins to {increase} the scatter by $10-40\%$ compared to the MZR at high redshift ($z>5$) in EAGLE.
The concept of an FMR is one that relies on minimizing scatter compared to the MZR, yet at the highest redshifts in EAGLE it achieves the opposite.
This is a clear failure of the strong FMR in EAGLE as well as a cautionary tale for interpreting future high-redshift FMR observations.

Finally, the right panel of Figure~\ref{fig:scatter_reduction} shows the ratio of the scatter of the weak FMR divided by the scatter of the strong FMR evaluated at each redshift ($\sigma_{\rm weak}/\sigma_{\rm strong}$).
The ratio $\sigma_{\rm weak}/\sigma_{\rm strong}$ is of particular interest as it provides a diagnostic for how well an assumed strong FMR characterises galaxies at higher redshift compared to their minimum scatter projection.
The ratio is unity at $z=0$ by construction, since the strong FMR assumes the $z=0$ $\alpha_{\rm min}$ value for all redshifts.
In Illustris and EAGLE, the scatter reduction of the weak FMR at $z\leq2$ is less than 5\%.
The relatively low decrease in the scatter in these two simulations implies that the strong FMR might approximately hold at these low redshifts (qualitatively consistent with previous observational findings; \citetalias{Mannucci_2010}; \citeauthor{Cresci_2019} \citeyear{Cresci_2019}). 
The scatter reduction at $z\geq3$, however, is $\gtrsim\!10\%$ for Illustris and EAGLE.
Both have monotonically decreasing ratios of scatter in the high-$z$ regime out to a 20\% decrease in Illustris and nearly 40\% in EAGLE.
On the other hand, the scatter reduction in TNG stays roughly constant at around 15\% at $z>0$.

Overall, the 10-40\% decrease in using a weak FMR indicates that high redshift galaxy populations are different from the low redshift systems.
The strong FMR does not effectively characterise these high redshift galaxies.
This marked shift in efficacy of the strong FMR further supports the idea there is some time evolution within the FMR in Illustris, TNG, and EAGLE.

It is worth noting how deceptive the lack of evolution in the strong FMR scatter reduction ratio (central panel of Figure~\ref{fig:scatter_reduction}) is for Illustris and TNG.
Looking at the reduction in scatter of the strong FMR by itself in these two simulations may lead one to conclude that a strong FMR holds -- the strong FMR does reduce scatter at all redshifts, even by a roughly constant amount.
Indeed it is remarkable that the strong FMR reduces the scatter by a similar amount at $z=0$ as $z=8$ despite ignoring a variation of a factor of $>2$ in $\alpha_{\rm min}$.
However, we emphasize that using the weak FMR significantly improves the characterisation of these galaxy populations, particularly at high redshift.

In summary, by determining $\alpha_{\rm min}$ at each redshift independently, we find that the scatter can be reduced an additional $\sim\!10-40\%$ compared to an assumed strong FMR.
We therefore conclude that the variations in $\alpha_{\rm min}$ are significant, particularly at high redshift.
The significant variations past $z\gtrsim3$ seem to imply that the strong FMR is not even a good approximation in the early universe in our simulations.

\section{Discussion}
\label{sec:discussion}

\subsection{What do variations in \texorpdfstring{$\alpha_{\rm min}$}{alpha min} mean?}\label{subsec:interpretation}

\begin{figure*}
    \centering
    \includegraphics[width=\linewidth]{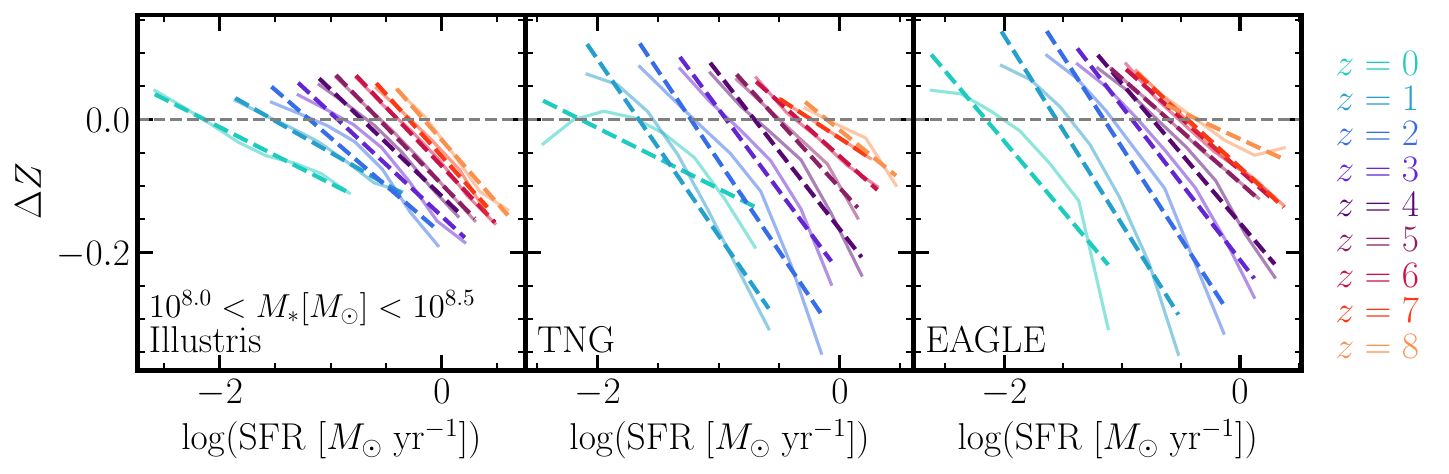}
    \caption{{\bf Offsets from the MZR as a function of SFR in a thin mass bin for $z=0-8$ in Illustris, TNG, and EAGLE.} The redshift evolution of the offsets from the MZR, $\Delta Z$, as a function of SFR for Illustris (left), TNG (centre), and EAGLE (right) for galaxies with stellar mass $10^{8.0} < M_*~[M_\odot] < 10^{8.5}$. The thin solid lines are the median offsets in fixed SFR bins of width 0.5 dex. The dashed lines are a linear regression of the medians. The different coloured lines are the different redshifts (left-to-right in an individual panel is $z=0-8$).  }
    \label{fig:dZdSFR}
\end{figure*}

The main idea of the FMR comes from the idea that in the MZR, at a fixed stellar mass, the metallicities and SFRs of galaxies are anti-correlated.
Using $\alpha_{\rm min}$ is an attempt to represent the strength of the correlation between metallicity and SFR; however, it does not actually explicitly tell us about that relationship.
Rather, $\alpha_{\rm min}$ values are tuned to minimize scatter.
It is therefore critical to develop an understanding of the strength of the correlation between metallicity and star formation rates that is not just a scatter-minimisation tool.
To build this understanding, we first take the FMR regression as defined in Section~\ref{subsec:FMR_param}:
\begin{equation}
    \label{eqn:FMR_form}
    Z = m\left(\log M_* - \alpha_{\rm min}\log {\rm SFR}\right) + b~,
\end{equation}
where $m$ is the slope of the regression and $b$ is the intercept\footnote{Note that we assume a linear regression here, but others (e.g., \citetalias{Mannucci_2010}) use a fourth-order regression. We show that our choice of first-order does not significantly impact our results in Appendix~\ref{appendix:polynomial}}.
By defining $\Delta Z = Z - \langle Z_{\rm MZR}\rangle$ (i.e., a galaxy's offset from the MZR is the metallicity of the galaxy subtracted from the MZR value at that galaxy's stellar mass) and take a thin mass bin, such that $\log M_*\approx C$, we can rearrange Equation~\ref{eqn:FMR_form} to obtain
\begin{equation}
    \label{eqn:DeltaZ}
    \Delta Z = m(-\alpha_{\rm min}\log {\rm SFR}) + b^\prime,
\end{equation}
where $b^\prime = b + mC - \langle Z_{\rm MZR}\rangle$.
Equation~\ref{eqn:DeltaZ} is a statement that, at fixed stellar mass, metallicity is anti-correlated with $\log {\rm SFR}$ by a factor of $m\alpha_{\rm min}$.
Here we have an explicit relationship between offsets from the MZR and the SFR of galaxies.
Conveniently, $\Delta Z$ is proportional to $-\alpha_{\rm min}$, our scatter minimisation parameter.
A key prediction here is that a strong FMR (no $\alpha_{\rm min}$ variations) should keep a constant relationship between offsets and SFR across time, whereas the weak FMR predicts a changing relationship\footnote{
We note, however, that while $\alpha_{\rm min}$ is {\it related} to the strength of the (anti-) correlation between $\Delta Z$ and SFR, in actuality, there is another scaling with the slope of the FMR ($m$)}.
The key advantage of considering FMR variations in $\Delta Z$-$\log {\rm SFR}$ space is in its interpretability.
We gain the same qualitative understanding by a smaller/larger $\alpha_{\rm min}$ value at high redshift, but in this framework it is more straight-forward to see how the relationship between metallicity and SFR changes with time.

To demonstrate $\Delta Z$'s scaling with SFR, we take a thin mass bin of width 0.5 dex ($10^{8.0} < M_*~[M_\odot] < 10^{8.5}$) and measure the offsets from the MZR as a function of SFR for all three simulations (see Figure~\ref{fig:dZdSFR}).
We note that these are qualitatively similar to the ``deviation plots'' of $\Delta Z$ and $\Delta$ specific SFR in \cite{Dave_2017}.
We determine the MZR from the median metallicity in fixed mass bins of width 0.05 dex. 
The offsets from the MZR, $\Delta Z$, are then generated by interpolating the MZR at each galaxy's stellar mass.

We find that, at all redshifts in each simulation, the offsets from the MZR are anti-correlated with SFR, as expected.
Furthermore, we find qualitative agreement between the slope of this anti-correlation and $\alpha_{\rm min}$ values.
In Illustris (left panel of Figure~\ref{fig:dZdSFR}), we find that the slope is shallow at $z=0$ and gets steeper with increasing redshift.
This behaviour is consistent with the $\alpha_{\rm min}$ variations seen in Illustris -- $\alpha_{\rm min}$ is small at $z=0$ in Illustris and increases with increasing redshift (see Figure~\ref{fig:alpha_redshift}).
We find that the $z=0$ slope in TNG is significantly weaker than the $z>0$ slopes (central panel of Figure~\ref{fig:dZdSFR}), consistent with the $\alpha_{\rm min}$ values from TNG.
It should be noted, however, that the $\alpha_{\rm min}$ values at $z\geq5$ are all the same, whereas the slopes of $\Delta Z$ versus SFR change slightly at $z\geq5$.
Finally, in EAGLE, we find that the slope of $\Delta Z$ versus SFR is steepest at $z=0$ and shallows with increasing redshift, again consistent with behaviour in $\alpha_{\rm min}$ as a function of redshift.
We emphasize that there is an additional term, $m$ (the slope of the FMR regression from Equation~\ref{eqn:FMR_form}), included in the slope of $\Delta Z$ versus SFR.
We therefore caution against too strong a comparison against $\alpha_{\rm min}$ and slopes in $\Delta Z$-SFR space.
Changes in the slope of the FMR may cause a change in the slope of $\Delta Z$ versus SFR.
$\Delta Z$ is only {\it proportional to} $\alpha_{\rm min}\log{\rm SFR}$.

\edit{
Furthermore, we note that slopes in $\Delta Z$-${\rm SFR}$ space have some dependence on the stellar mass bin chosen.
This behaviour is consistent with more recent parameterisations of the FMR \citep[e.g.,][]{Curti_2020} that find that the MZR turnover mass has some dependence on SFR.
% We find that derived $\alpha_{\rm min}$ values have some deviation when measured in thin stellar mass bins.
While an interesting area of future exploration, the stellar mass dependence of these slopes (and $\alpha_{\rm min}$) is beyond the scope of this work.
We therefore caution that the model presented here neglects variations in the role of SFR as a function of stellar mass.
However, since Equation~\ref{eqn:DeltaZ} explicitly assumes a thin mass bin, the simple model presented here should be relatively robust to stellar mass variations since the $\Delta Z$-${\rm SFR}$ space slopes are not {\it highly} sensitive to mass (which we find to be predominately be the case in each simulation at each redshift).
}

In summary, the slope between offsets from the MZR and SFR offers a more straight-forward way to understand the (potential) evolution in the relationship between metallicity and SFR suggested by $\alpha_{\rm min}$ variations.

\begin{figure*}
    \centering
    \includegraphics[width=0.95\linewidth]{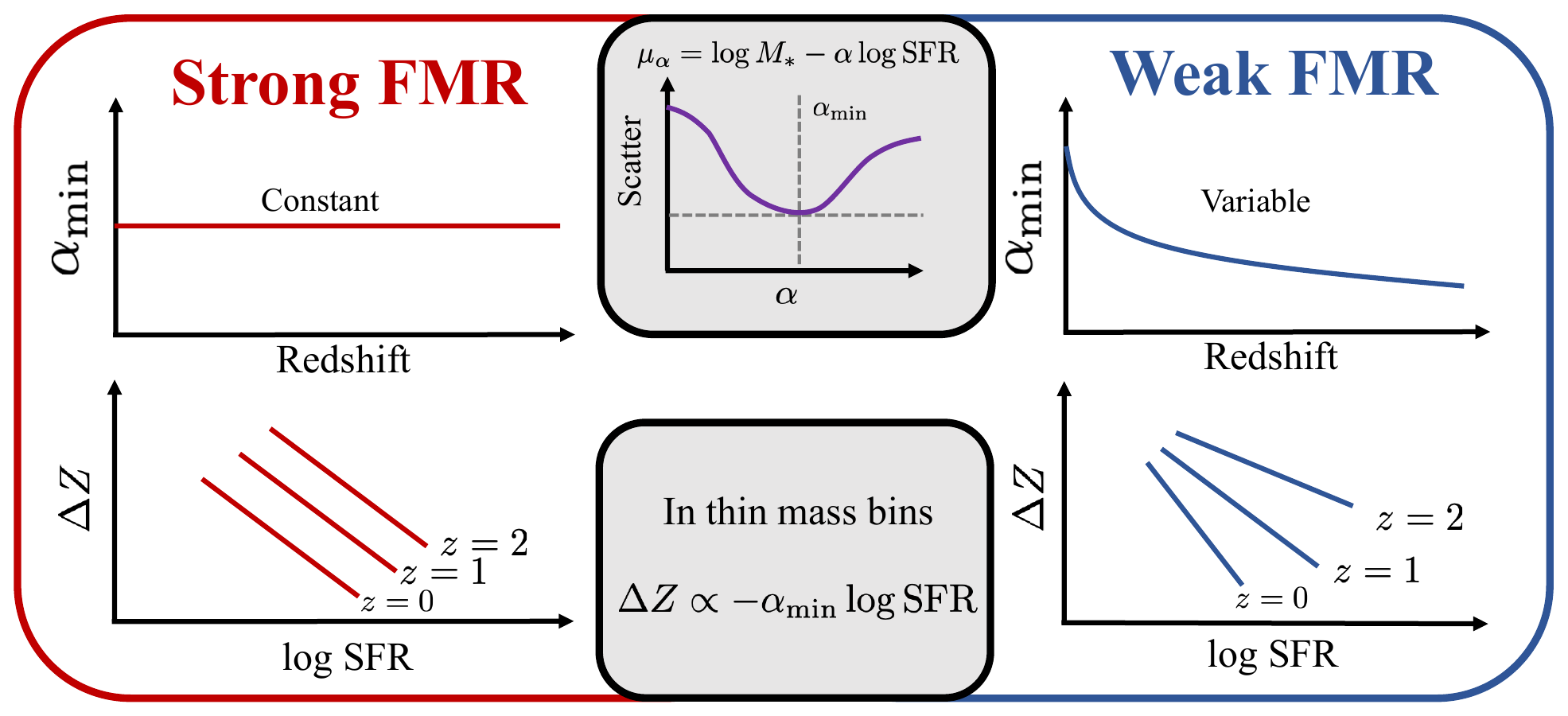}
    \caption{{\bf Summary of Key Points.} The strong FMR (left, red) is where $\alpha_{\rm min}$, a parameter tuned to minimise scatter about the MZR, is constant as a function of redshift. Consequently, in thin mass bins, the offsets from the MZR, $\Delta Z$, as a function of (log) SFR have roughly the same slope at all redshifts (although there is a dependence on the slope of the FMR, see Section~\ref{subsec:interpretation} for more details). The weak FMR (right, blue) is where $\alpha_{\rm min}$ varies as a function of redshift. In this scenario, the individual redshift's have different strengths of correlations between offsets from the MZR and (log) SFR. }
    \label{fig:summary_fig}
\end{figure*}

\subsection{Advantages and challenges of a weak FMR Framework}\label{subsec:evaluting_FMR}

The key advantage of fitting each redshift independently is to more effectively minimize the scatter.
The weak FMR gives us a clear-cut metric for the strength of the MZR's secondary dependence on SFR arises that is completely independent of the evolution of the normalisation of the MZR.
Independence from other redshift populations removes the possibility of conflating evolution of the normalisation of the MZR with evolution of the scatter.
By using a $z=0$ derived $\alpha_{\rm min}$ at all redshifts (i.e., strong FMR) we suppress any potential variation of $\alpha_{\rm min}$ as a function of redshift.
As a consequence, the strong FMR does not optimally reduce scatter across redshift (discussed in more detail in Section~\ref{subsubsec:discussion_alphamin}).
Using a weak FMR assumption therefore allows a more careful examination for the extent to which the observed FMR has variations.
% If evolution exist and $\alpha_{\rm min}$ does not vary, it can be concluded that evolution {\it must} be driven by the evolution of the normalisation of the MZR.

A challenge of performing a similar analysis in observations is the amount of data available.
For example, lower redshift galaxy populations are well sampled (e.g., \citetalias{Mannucci_2010} use 141,825 SDSS $z\sim0$ galaxies), but at higher redshift sampling becomes more difficult (e.g., the \citeauthor{Nakajima_2023} \citeyear{Nakajima_2023} and \citeauthor{Curti_2023} \citeyear{Curti_2023} analyses use less than 200 objects spanning a wider redshift range of $z=3-10$).
It is possible that subtle changes can be measured at lower redshifts (see \citeauthor{Pistis_2023} \citeyear{Pistis_2023} for a potential detection of MZR scatter variations at $z\sim0.63$); however, the most significant scatter reduction happens in the high redshift ($z\gtrsim3$) populations (Figures~\ref{fig:scatter_reduction}).
More complete samples of galaxy populations at these early times with, e.g., JWST are therefore required in order to undergo any weak FMR-style analysis to detect significant deviations from the $z=0$ $\alpha_{\rm min}$ values.
Moreover, a redshift-complete sample would be limited by our understanding of metallicity in the high redshift universe.
Recently, work has been done to obtain reliable metallicity diagnostics at $z>4$ using \JWST{}/NIRSpec \citep[e.g.,][]{Nakajima_2023,Shapley_2023,Sanders_2024}.
However, more complete galaxy samples are required, particularly at the low metallicities seen at this epoch, to fully characterise these diagnostics.
As such, it is currently difficult to ensure that $\alpha_{\rm min}$ values determined observationally are fair comparisons across the broad redshift range examined in this work.

\subsection{Dependence on small scale physics implementations}\label{subsec:subgrid_physics}

We find that Illustris, TNG, and EAGLE have weak FMRs.
The $\alpha_{\rm min}$ values are not the same, nor do they evolve in the same fashion, in the different models, however.
The value of $\alpha_{\rm min}$ at any given redshift is a complicated by-product of a number of different physical processes.
\edit{
We therefore caution the reader against drawing conclusions on $\alpha_{\rm min}$ (and the evolution thereof) from the aggregation of the three individual models.
}
While we have some qualitative understanding of how $\alpha_{\rm min}$ is set (or changed), the exact mechanisms \edit{setting} $\alpha_{\rm min}$ are not entirely clear in detail.

What is clear is that $\alpha_{\rm min}$ is sensitive to the physics driving galaxy evolution.
For example, in Section~\ref{subsec:alpha_redshift}, we attributed the lowered $\alpha_{\rm min}$ values in TNG at $z=0$ to the redshift-dependent wind prescription in the TNG model (as mentioned in Section~\ref{subsec:alpha_redshift}).
Through this example, the sensitivity of $\alpha_{\rm min}$ to the input physics within the simulation models becomes clear.
The redshift-dependent winds in TNG work to increase wind velocities at low redshift which suppresses star formation.
This star formation suppression likely plays a significant role in the overall decrease of $\alpha_{\rm min}$ seen at low redshifts in TNG.
\edit{
We therefore propose the evolution (or lack thereof) in the scatter about the MZR as a testable prediction to constrain the physical models of the simulations.
}

All three models examined here rely on effective equation of state sub-grid models for the dense, unresolved ISM (\citeauthor{Springel_Hernquist_2003} \citeyear{Springel_Hernquist_2003} for Illustris/TNG and \citeauthor{Schaye_DallaVechhia_2008} \citeyear{Schaye_DallaVechhia_2008} for EAGLE).
In recent years, however, high-resolution simulation modelling has begun to directly resolve the sites of star formation (e.g., Feedback In Realisitc Environments model; \citeauthor{Hopkins_2014} \citeyear{Hopkins_2014}).
The stellar feedback in such simulations is much burstier than in the models presented here.
We believe that bursty stellar feedback events should suppress $\alpha_{\rm min}$ values compared to Illustris, TNG, and EAGLE.
Subgrid pressurization lends support to the ISM that is not coupled to star formation, and therefore blunts rapid variations in star formation and stellar feedback. 
Models without subgrid pressurization (like FIRE) do not have this source of ISM support, and therefore exhibit more rapid (i.e., bursty) variations in star formation and stellar feedback.
Bursts may therefore curtail the effectiveness of star formation rates in regulating the gas-phase metallicity of a galaxy.
Therefore the redshift variations in $\alpha_{\rm min}$ may be able to provide constraining power on the extent to which galaxies' feedback is more bursty or smooth.
Although it should be noted that even within these smooth feedback models there is some disparity.

\section{Conclusions}
\label{sec:conclusion}

We select central star forming galaxies with stellar mass $8.0 < \log( M_*~[M_\odot] ) < 12.0$ with gas mass $\log (M_{\rm gas}~[M_\odot]) > 8.5$ from $z=0-8$ in the cosmological simulations Illustris, IllustrisTNG, and EAGLE.
We investigate the extent to which the \citetalias{Mannucci_2010} parameterisation (see Equation~\ref{eqn:mu_M10}; $\muZR$) of the fundamental metallicity relation (FMR; Equation~\ref{eqn:mu_M10}) holds.
The parameter of merit in the $\muZR$ is $\alpha_{\rm min}$, which is a parameter tuned to minimize scatter in the relation.
Physically, $\alpha_{\rm min}$ sets a projection direction of the mass-metallicity-SFR space to a 2D space with minimal scatter.
Many observational studies have claimed that this projection direction does not evolve with redshift \citep{Mannucci_2010,Cresci_2019}.

We discuss a new framework in which to examine the $\muZR$ as a superset of the MZR ($\alpha=0$) and FMR ($\alpha\neq0$).
We further define both a strong and weak FMR.
A strong FMR indicates that $\alpha_{\rm min}$ is constant as a function of redshift.
Conversely, the weak FMR is where $\alpha_{\rm min}$ varies with redshift (see Figure~\ref{fig:general_muZR} for complete illustrated relationship of $\muZR$).
More generally, the strong FMR states the the \citetalias{Mannucci_2010} parameterisation can describe both the scatter and noramlisation of the MZR at the same time.

Our conclusions are as follows:

\begin{itemize}
    \item We find that $\alpha_{\rm min}\neq0$ for all redshifts in Illustris, TNG, and EAGLE. This shows that there is an FMR in each of these simulations. We note, however, that the uncertainty in $\alpha_{\rm min}$ in TNG at $z=0$ includes $\alpha_{\rm min}=0.0$. We attribute this to the increased suppression of low redshift star formation in the TNG model.
    \item Furthermore, we find that there is non-negligible evolution in $\alpha_{\rm min}$ as a function of redshift (Figure~\ref{fig:alpha_redshift}).
    This result suggests that the FMR in Illustris, TNG, and EAGLE is a weak FMR.
    \item We find that the weak FMR ($\alpha_{\rm min}$ determined at each redshift independently) consistently reduces scatter around $10-30\%$ compared to the the MZR (left panel of Figure~\ref{fig:scatter_reduction}).
    The strong FMR also reduces the scatter compared to the MZR, albeit to a lesser extent than the weak FMR.
    At high-$z$ in EAGLE, however, using the strong FMR actually {\it increases} scatter compared to the MZR (centre panel of Figure~\ref{fig:scatter_reduction}).
    Overall, we find that at $z\gtrsim3$ fitting galaxies with a weak FMR can reduce scatter $\sim5-40\%$ more than using the strong FMR (right panel of Figure~\ref{fig:scatter_reduction}).
    \item We suggest that the interpretation of $\alpha_{\rm min}$ variations is more well-understood in the context of the slope of $\Delta Z$ (offsets from the MZR) as a function of $\log {\rm SFR}$ (see Figure~\ref{fig:dZdSFR}). In this context, the weak FMR suggest that the relationship between metallicity and SFR changes through cosmic time, whereas the strong FMR suggests that it does not change. We also show that the slope in $\Delta Z\!-\!\log {\rm SFR}$ space is proportional to $\alpha_{\rm min}$ (see Equation~\ref{eqn:DeltaZ}).
\end{itemize}

Obtaining one relationship that describes the metal evolution of all galaxies across time is an ambitious goal.
It is worth appreciating how reasonably well a simple linear combination of two parameters can begin to achieve that goal at low redshift.
Yet it is not perfect.
To begin to rectify this, we develop a substantial overhaul to the current FMR paradigm (summarized in Figure~\ref{fig:summary_fig}).
The results from this work show that Illustris, TNG, and EAGLE indicate deviations from the strong FMR.
It is presently unclear whether the same is true in observations.
Understanding whether the FMR in observations is weak or strong will aid in being able to understand the recent \JWST{} observations suggesting high redshift FMR evolution.

\section*{Acknowledgements}

\edit{We thank the anonymous referee for their thoughtful comments and feedback on improving the quality of this manuscript.}
AMG acknowledges Carol, Kate, and Kelly Garcia for assistance in the design of Figures~\ref{fig:general_muZR}~and~\ref{fig:summary_fig}.
We acknowledge the Virgo Consortium for making their simulation data available. The EAGLE simulations were performed using the DiRAC-2 facility at Durham, managed by the ICC, and the PRACE facility Curie based in France at TGCC, CEA, Bruy\`eresle-Ch\^atel.

AMG and PT acknowledge support from NSF-AST 2346977.
KG is supported by the Australian Research Council through the Discovery Early Career Researcher Award (DECRA) Fellowship (project number DE220100766) funded by the Australian Government. 
KG is supported by the Australian Research Council Centre of Excellence for All Sky Astrophysics in 3 Dimensions (ASTRO~3D), through project number CE170100013. 
RJW acknowledges support from the European Research Council via ERC Consolidator Grant KETJU (no. 818930)

%%%%%%%%%%%%%%%%%%%%%%%%%%%%%%%%%%%%%%%%%%%%%%%%%%
\section*{Data Availability}

The reduced data products, analysis scripts, and figures are all available publicly at \href{https://github.com/AlexGarcia623/Does-the-FMR-evolve-Simulations/}{https://github.com/AlexGarcia623/Does-the-FMR-evolve-Simulations/}.
Data from Illustris and IllustrisTNG is publicly available on each project's respective website. 
Illustris: \href{https://www.illustris-project.org/data/}{https://www.illustris-project.org/data/} and IllustrisTNG: \href{https://www.tng-project.org/data/}{https://www.tng-project.org/data/}.
Similarly, data products from the EAGLE simulations are available for public download via the Virgo consortium’s website: \href{https://icc.dur.ac.uk/Eagle/database.php}{https://icc.dur.ac.uk/Eagle/database.php}

%%%%%%%%%%%%%%%%%%%% REFERENCES %%%%%%%%%%%%%%%%%%

% The best way to enter references is to use BibTeX:

\bibliographystyle{mnras}
\bibliography{bibliography} % if your bibtex file is called example.bib

%%%%%%%%%%%%%%%%%%%%%%%%%%%%%%%%%%%%%%%%%%%%%%%%%%

%%%%%%%%%%%%%%%%% APPENDICES %%%%%%%%%%%%%%%%%%%%%

\appendix

\section{(Lack of) Dependence on Specific star formation main sequence}\label{appendix:sSFMS_cuts}

\begin{figure}
    \centering
    \includegraphics[width=\linewidth]{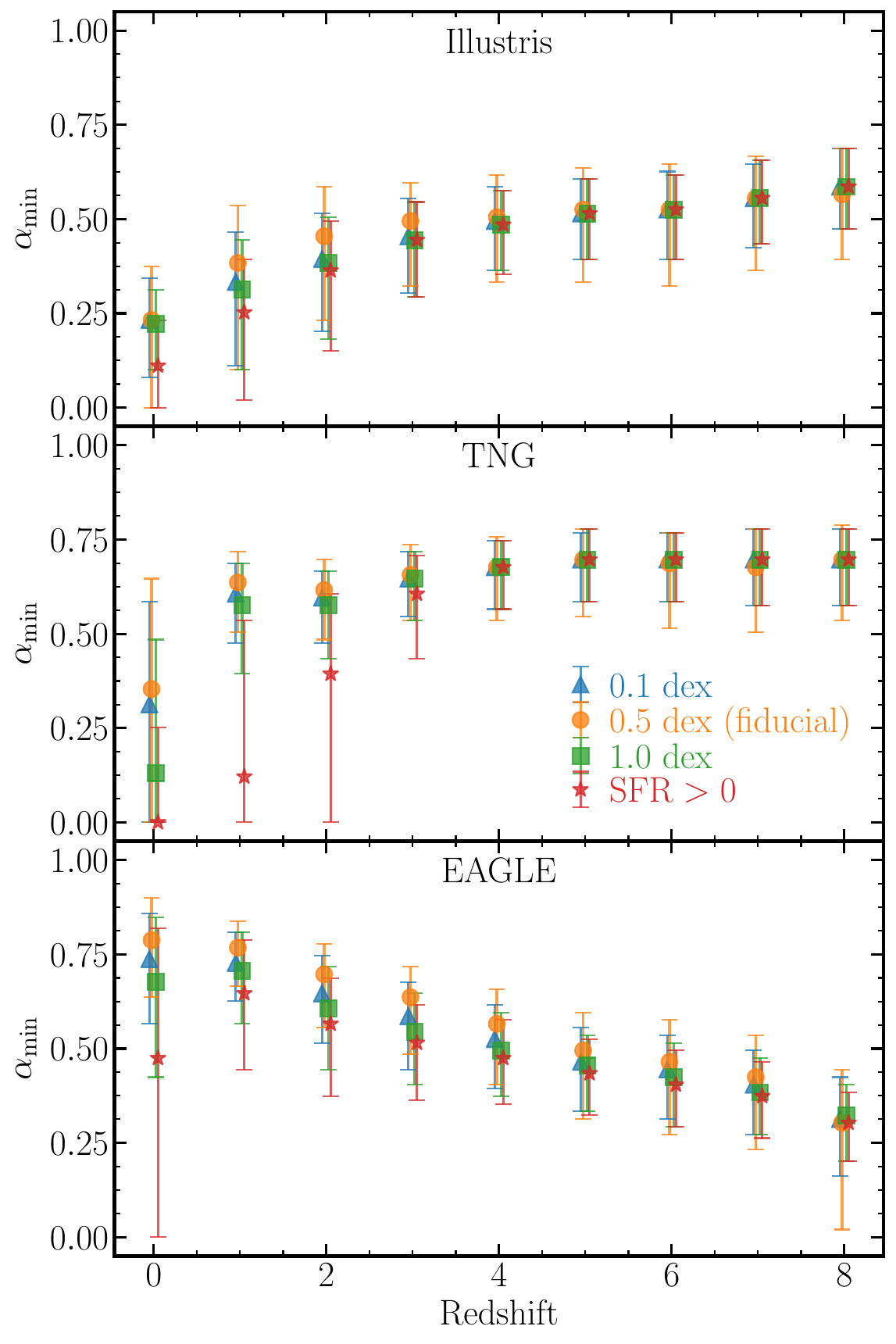}
    \caption{Determination of $\alpha_{\rm min}$ as a function of redshift for Illustris (top), TNG (centre), and EAGLE (bottom) for the four different sSFMS variations.}
    \label{fig:sSFMS_comparison}
\end{figure}

Part of our galaxy selection criteria includes selecting star forming galaxies (see Section~\ref{subsec:galaxy_selection} for full details).
Our method of selecting these galaxies uses a specific star formation main sequence cut.
The sSFMS selection includes a cut excluding galaxies 0.5 dex below the median relation.
In this appendix, we consider three additional variations on this cut (following from \citeauthor{Garcia_2024} \citeyear{Garcia_2024}): (i) a more restrictive cut of all galaxies 0.1 dex below the median relation, (ii) a less restrictive cut of all galaxies 1.0 dex below the median relation, and (iii) a very liberal cut of all galaxies with non-zero SFRs.
We show the resultant $\alpha_{\rm min}$ values from these cuts in Figure~\ref{fig:sSFMS_comparison}.
The uncertainty bars on $\alpha_{\rm min}$ overlap for all redshift bins in all three analysed simulations with all four cuts.
However, there are three cases in which the derived $\alpha_{\rm min}$ value itself varies significantly for the SFR $>0$ cut: TNG $z=1$ and $2$ as well as EAGLE $z=0$.
In these three cases, $\alpha_{\rm min}$ is significantly offset from the errorbars of at least two of the three other cuts.
The uncertainties using the SFR $>0$ cut in these three cases are quite large.
This suggests that the overall change in scatter when using the SFR $>0$ cut versus another cut is marginal.
We therefore conclude that, while the derived $\alpha_{\rm min}$ value may change, these changes have no qualitative bearing on the results presented in this work.

% \subsection{Minimum Gas Mass}

% \begin{figure}
%     \centering
%     \includegraphics[width=\linewidth]{FigureA2.pdf}
%     \caption{Determination of $\alpha_{\rm min}$ as a function of redshift for Illustris (top), TNG (centre), and EAGLE (bottom) for three different minimum gas mass variations.}
%     \label{fig:Mgas_comparison}
% \end{figure}

% \subsection{Maximum Stellar Mass}

% \begin{figure}
%     \centering
%     \includegraphics[width=\linewidth]{FigureA3.pdf}
%     \caption{Determination of $\alpha_{\rm min}$ as a function of redshift for Illustris (top), TNG (centre), and EAGLE (bottom) for three maximum stellar mass variations.}
%     \label{fig:Mstar_comparison}
% \end{figure}

\section{Higher order polynomial fit}\label{appendix:polynomial}

\begin{figure}
    \centering
    \includegraphics[width=\linewidth]{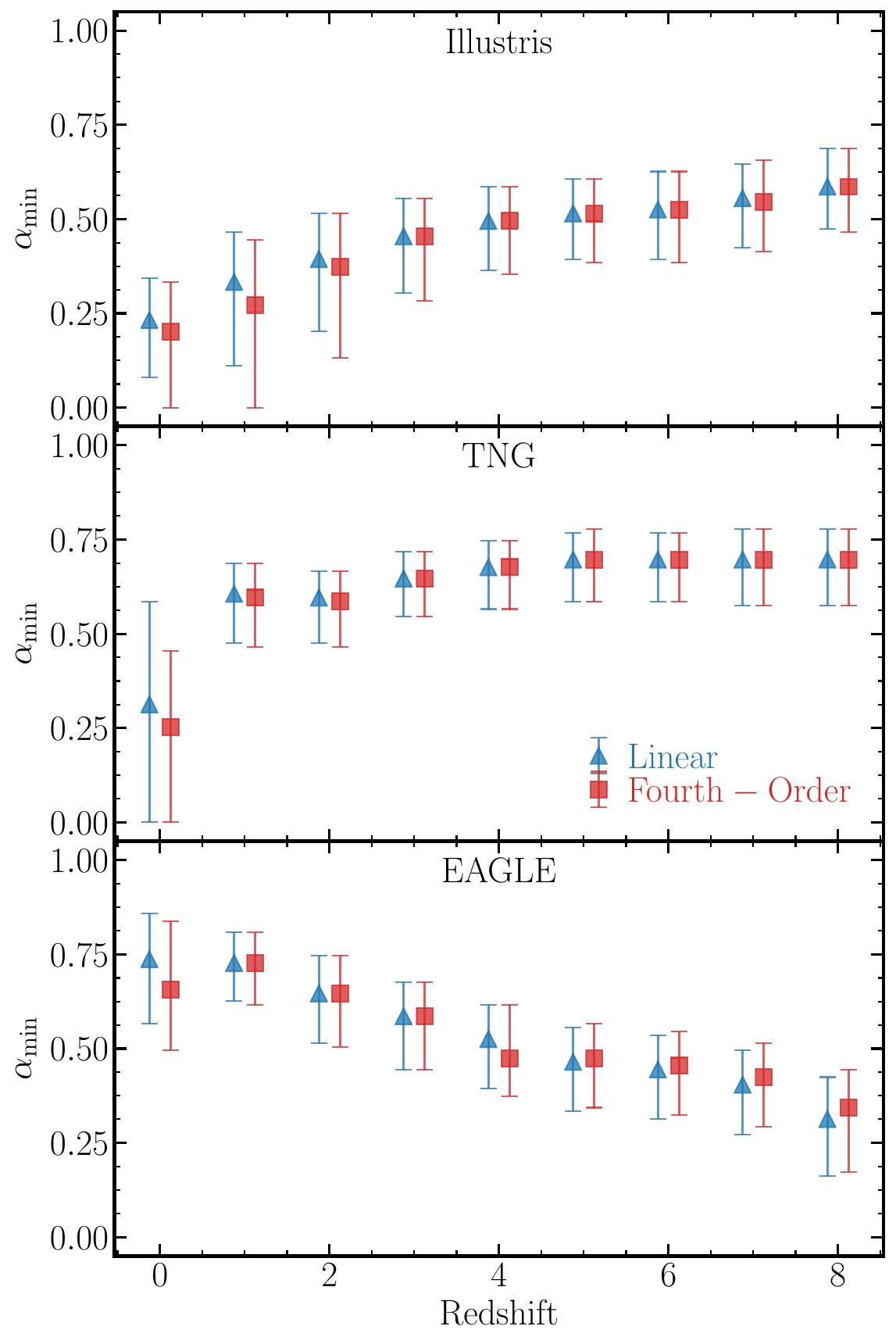}
    \caption{Determination of $\alpha_{\rm min}$ as a function of redshift for Illustris (top), TNG (centre), and EAGLE (bottom) using a linear regression (blue triangles) and fourth-order polynomial regression (red squares).}
    \label{fig:linearV4th}
\end{figure}

\citetalias{Mannucci_2010} determined residuals in the scatter about a fourth-order polynomial instead of a linear regression.
This practice is not consistent through all works using the $\mu_{\alpha}$ 2D projection of the FMR, however.
For example, recent \JWST{} observational papers \citep[e.g.,][]{Nakajima_2023,Langeroodi_2023} adopt a linear regression definition of the FMR from \cite{Andrews_Martini_2013}.
We show that using a linear regression does not significantly change the projection of least scatter in Illustris, TNG, and EAGLE in Figure~\ref{fig:linearV4th}.

%%%%%%%%%%%%%%%%%%%%%%%%%%%%%%%%%%%%%%%%%%%%%%%%%%

% Don't change these lines
\bsp	% typesetting comment
\label{lastpage}
\end{document}

%% file: figures/Figure1.tikz
\newcommand{\sSFR}{ {\rm sSFR}^{-1}\!\!\!-\!{\rm ZR} }
\def\alpMZR{\alpha=0.0}
\def\alpFMR{\alpha\neq0.0}

\definecolor{Mycolor1}{HTML}{A93F53}
\definecolor{Mycolor2}{HTML}{3DA6A1}
\definecolor{Mycolor3}{HTML}{FEFB39}

\def\colorOne{Mycolor1!60!red!60}
\def\colorTwo{Mycolor2!60!white}
\def\colorThree{Mycolor3!70!white}

\begin{tikzpicture}
    \foreach \angle\name\alpVal\x in {-40/MZR/$\alpMZR$/-1,40/FMR/$\alpFMR$/1} {
        \begin{scope}[rotate around={\angle:(0,0)}]
            \draw[->,very thick] (0,0) -- (0,-2.1);
            \coordinate (\name) at (0,-2.75);
        \end{scope}
        \draw[fill=\colorTwo, thick] (\name) ellipse (0.8 and 0.35) node[] {\name};
        \node[fill=white] at (\x,-1) {\alpVal};
    }
    % \foreach \angle\name\alpVal\x\label\col in {-40/sSFR/$\alpsSFR$/0.25/$\sSFR$/\colorTwo,40/FMR/$\alpFMR$/-0.25/FMR/\colorTwo} {
    %     \begin{scope}[rotate around={\angle:(0,0)}]
    %         \draw[->,very thick] (0,0) -- (0,-2);
    %         \coordinate (\name) at (0,-2.5);
    %         \node[fill=white] at (\x,-3) {\alpVal};
    %     \end{scope}
        
    %     \draw[fill=\col, thick] (\name) ellipse (0.8 and 0.35) node[] {\label};
    % } 
    \draw[ fill=\colorOne, thick ] (0,0) ellipse (1 and 0.5) node[] {$\muZR$}; 
    \node[] at (0,0.75) {\large $\mu_{\alpha} = \log M_* - \alpha\log {\rm SFR}$};

    \foreach \angle\name\label in {-40/LFMR/Weak,40/aFMR/Strong} {
        \begin{scope}[rotate around={\angle:(FMR)}]
            \draw[->,very thick] ($(FMR) - (0,0.5)$) to ($(FMR) - (0,2)$);
             \coordinate (\name) at ($(FMR) - (0,2.5)$);
        \end{scope}
        \draw[fill=\colorThree, thick] (\name) ellipse (0.8 and 0.35) node[] {\label};
    }

    % \draw[->,very thick] ($(RSZR) - (0,0.5)$) to[in=70,out=-70] ($(aFMR) - (0.1,-0.4)$);  \node[anchor=west, rotate=-90, draw=gray, thick] at ($(RSZR) - (-0.85,0.5)$) {No $\alpha_{\rm min}$ variation};

    % \node[] at ($(FMR) - (0,1.25)$) {$\alpha_{\rm min}$ determined at};
    \node[fill=white] at ($(LFMR) + (0.66,1)$) {$\alpha=\alpha(z)$};
    \node[fill=white] at ($(aFMR) + (-0.66,1)$) {$\alpha=C$};

    % \draw[->,very thick] ($(LFMR) - (-0.25,0.75)$) to[out=-38,in=-165] ($(aFMR) - (0.8,0.25)$);
    % \draw[thick,color=gray] ($(FMR) - (3,2.8)$) rectangle ($(FMR) - (-1.5,3.7)$);
    % \node[] at ($(FMR) - (0.75,3)$) {Slope/intercept same at $z\approx0$ as all $z$};
    % \node[] at ($(FMR) - (0.75,3.25)$) {\&};
    % \node[] at ($(FMR) - (0.75,3.5)$) {$\alpha_{{\rm min}} (z=0) = \alpha_{{\rm min}}({\rm all~}z)$};

    % \draw[->, very thick] ($(aFMR) - (-0.8,0.3)$) to [out=-20,in=60] ($(aFMR) - (-0.8,2.0)$) to[out=-120,in=0] ($(FMR) - (-1,4.5)$);
    % \draw[fill=yellow!70!white, thick] ($(FMR) - (0,4.5)$) ellipse (0.8 and 0.35) node[] {(ri)FMR} coordinate (riFMR);
    % \node[draw=gray,thick] at ($(riFMR) - (0,0.6)$) {Local $=$ Global};
    % \draw[dashed, very thick] ( -1.5,-3.5 ) -- ( 2.25 , 0.75 );
    % \node[] at (3.1,1) {\large Literature};
    % \node[] at (3.1,0.6) {\large ``FMR''};
\end{tikzpicture}

%% file: tables/table1.tex
\begin{table}
    \centering
    \begin{tabular}{lx{0.15\linewidth}x{0.15\linewidth}x{0.15\linewidth}}
         \toprule
          & {\bf Illustris} & {\bf TNG} & {\bf EAGLE} \\\midrule
         $z=0$ & 0.23$_{0.08}^{0.34}$ & 0.31$_{0.00}^{0.59}$ & 0.74$_{0.57}^{0.86}$\\[0.5em]
         $z=1$ & 0.33$_{0.11}^{0.46}$ & 0.61$_{0.47}^{0.69}$ & 0.73$_{0.63}^{0.81}$\\[0.5em]
         $z=2$ & 0.39$_{0.20}^{0.52}$ & 0.60$_{0.47}^{0.67}$ & 0.65$_{0.52}^{0.75}$\\[0.5em]
         $z=3$ & 0.45$_{0.30}^{0.56}$ & 0.65$_{0.55}^{0.72}$ & 0.59$_{0.44}^{0.68}$\\[0.5em]
         $z=4$ & 0.49$_{0.36}^{0.59}$ & 0.68$_{0.57}^{0.75}$ & 0.53$_{0.39}^{0.62}$\\[0.5em]
         $z=5$ & 0.52$_{0.39}^{0.61}$ & 0.70$_{0.59}^{0.77}$ & 0.46$_{0.33}^{0.56}$\\[0.5em]
         $z=6$ & 0.53$_{0.39}^{0.63}$ & 0.70$_{0.59}^{0.77}$ & 0.44$_{0.31}^{0.54}$\\[0.5em]
         $z=7$ & 0.56$_{0.42}^{0.65}$ & 0.70$_{0.58}^{0.78}$ & 0.40$_{0.27}^{0.49}$\\[0.5em]
         $z=8$ & 0.59$_{0.47}^{0.69}$ & 0.70$_{0.58}^{0.78}$ & 0.31$_{0.16}^{0.42}$\\\bottomrule
         % all $z$ & 0.26 & 0.54 & 0.71 \\\bottomrule
    \end{tabular}
    \caption{{\bf All $\alpha_{\rm min}$ values at $z=0-8$ for Illustris, TNG, and EAGLE.} These $\alpha_{\rm min}$ values are determined at each redshift individually. The superscripts are the upper limits of the uncertainty while the subscripts are the lower limits. We show these values in Figure~\ref{fig:alpha_redshift}.}
    \label{tab:alphas}
\end{table}